\documentclass[12pt]{iopart}

\usepackage{iopams}  
\usepackage{graphicx}
\usepackage{xcolor}

\begin{document}

\title[Shock boundary oscillations]{Oscillations of subcritical fast magnetosonic shock boundaries caused by shock reformation}

\author{M. E. Dieckmann}
\address{Dept. of Science and Technology (ITN), Link\"oping University, Campus Norrk\"oping, SE-60174 Norrk\"oping, Sweden.}
\ead{mark.e.dieckmann@liu.se}

\author{A.~Bret}
\address{ETSI Industriales, Universidad de Castilla-La Mancha, 13071, Ciudad Real, Spain.}
\address{Instituto de Investigaciones Energéticas y Aplicaciones Industriales, Campus Universitario de Ciudad Real, 13071, Ciudad Real, Spain.}

\author{D.~Folini}
\address{University of Lyon, ENS de Lyon, Univ Lyon 1, CNRS, Centre de Recherche Astrophysique de Lyon UMR5574, F-69230, Saint-Genis-Laval, France.}

\author{R.~Walder}
\address{University of Lyon, ENS de Lyon, Univ Lyon 1, CNRS, Centre de Recherche Astrophysique de Lyon UMR5574, F-69230, Saint-Genis-Laval, France.}

\vspace{10pt}
\begin{indented}
\item[]July 2024
\end{indented}

\begin{abstract}
The evolution of a deformed subcritical fast magnetosonic shock front is compared between two two-dimensional PIC simulations with different orientations of the magnetic field relative to the simulation box. All other initial and simulation conditions are kept identical. Shock boundary oscillations are observed in the simulation where the magnetic field direction is resolved. This oscillation is caused by the reformation of the shock front. One part of the front acts as a shock, while the other functions as a magnetic piston, with both halves changing their states in antiphase. The oscillation period corresponds to the time required for one shock wave to grow as the other collapses. In contrast, the corrugated fast magnetosonic shock does not oscillate in the second simulation, where the magnetic field is oriented out of the simulation plane. This dependence on magnetic field orientation suggests that the shock oscillation is induced by magnetic tension, which is only effective in the first simulation. In both simulations, the shock perturbation does not grow over time, indicating that the shocks are stable. The potential relevance of these findings for the Alfvénic oscillations of the supercritical Earth's bow shock, detected by the MMS multi-spacecraft mission, is also discussed. 
\end{abstract}

%
\vspace{2pc}
\noindent{\it Keywords}: fast magnetosonic shock, PIC simulations, shock boundary oscillations

\submitto{\PPCF}
%
\maketitle
%
%

\section{Introduction}

Subcritical shocks in unmagnetized~\cite{Forslund1970,Bardotti1970} or magnetized~\cite{Forslund1971} collisionless plasma are sustained by the change in the electric potential across their boundary, which is caused by the net diffusion of electrons from the denser downstream plasma into the upstream plasma. In the rest frame of the shock, the positive electric potential of the downstream plasma causes the inflowing upstream ions to slow down and compress. The velocity spread of the ions is also increased, raising their temperature along the shock normal. This occurs because ions with large velocities along the shock normal experience a smaller relative change in kinetic energy in this direction than slower ions when crossing the shock. Ions are heated only along the shock normal, implying that the ions can only be compressed to twice their upstream density~\cite{Livadiotis2015}. The resulting small jump in the electric potential limits the maximum speed of stable shocks. 

The wave mediating the shock is determined by whether or not a magnetic field is present. In unmagnetized plasma, it is the ion-acoustic mode, whereas in magnetized plasma with a magnetic field orthogonal to the shock normal, the fast magnetosonic mode is involved. In the limit of low frequencies, the phase and group velocities of ion-acoustic and fast magnetosonic modes are constant. Dispersionless waves with a low amplitude cannot form a stable shock (see, for example, Ref.\cite{Bret2022}). Waves with a large amplitude steepen~\cite{Dawson1959,Shukla2004}, transferring wave power to increasingly large wavenumbers and frequencies. Eventually, the wave frequencies approach resonance, reducing both the phase and group velocities. Waves with a short wavelength fall behind thereby halting a further steepening of the wave front as it was demonstrated in a particle-in-cell (PIC) simulation of a fast magnetosonic shock~\cite{Dieckmann2017}. Therefore, a stable subcritical shock is a dispersive wave packet that moves through the downstream plasma and faces the upstream plasma at its front. Its electric field along the shock normal causes oscillations in the ion density and mean velocity, with the largest changes in ion velocity occurring at the front of the wave, bridging the velocity gap across the shock.

In the upstream frame of reference, subcritical fast magnetosonic shocks, on which we focus here, can travel at speeds 2-3 times larger than the fast magnetosonic speed~\cite{Marshall1955,Edmiston1984}. A thermal spread of upstream ions results in some ions being unable to cross the shock, causing them to be reflected back upstream. The directed flow energy of this ion beam is then released through collisionless instabilities. Subcritical shocks reflect only a small fraction of ions, keeping upstream instabilities weak. 
Supercritical shocks~\cite{Quest1986,Lembege1992,Lowe2003,MTSI,Marcowith2016,Dimmock2019}, which move at larger Mach numbers, are non-stationary and they reflect a substantial fraction of the inflowing ions driving strong instabilities upstream. These instabilities modulate the density and magnetic field of the upstream plasma, leading to the corrugation of the shock front. Hybrid simulations, which employ a kinetic plasma model for ions and represent electrons as an inertialess fluid, have shown that the boundary of a supercritical shock begins to oscillate in the form of Alfv\'en waves as it propagates through the nonuniform plasma~\cite{Lowe2003}. Waves propagating along the magnetic field of a quasi-perpendicular supercritical shock boundary were also found at the Earth's bow shock~\cite{Johlander2016}. In both cases, the shock oscillations could not be examined in isolation, as they were coupled with upstream perturbations.

PIC simulations were performed, employing a kinetic model for electrons and ions, in which a subcritical and planar fast magnetosonic shock propagated across a perturbation layer. In this layer, the number density of mobile ions varied sinusoidally along the direction of the shock boundary. As the shock crossed this layer, its boundary was deformed in the direction of the shock's average normal. This deformation varied sinusoidally, with an amplitude small compared to the deformation's wavelength. Upon entering the unperturbed plasma, oscillations were observed in the two-dimensional simulation that resolved the direction of the magnetic field numerically~\cite{Dieckmann2023}. The oscillation frequency was just below the lower-hybrid frequency, which is also the frequency of the wave packet that mediates the shock. A second shock in the same simulation, which propagated into a uniform ambient plasma with the same composition and magnetic field orientation serving as reference shock, moved at the same speed as the perturbed shock and remained planar until the simulation's end. In a second simulation, detailed in~\cite{Dieckmann2024}, where the magnetic field pointed out of the two-dimensional simulation box, the perturbation of the shock front was non-oscillatory and weakly damped. The reference shock in this second simulation did not remain planar. The drift of electrons in the density gradient just behind of the shock's density overshoot led to the growth of lower-hybrid waves, which deformed the shock boundary and led to a nonuniform flow of ions across the shock. However, the shock remained stable. It was concluded that magnetic tension is essential for the propagation of waves along the shock's magnetic field. The mechanism responsible for the boundary oscillations observed in the first simulation was, however, not identified unambigously. Here we find by a direct comparison of the data from both simulations that cyclic reformation of the shock is responsible for the shock boundary oscillations.

Our paper is organized as follows. Section~2 discusses the simulation setup used to deform the shock boundary and dispersive properties of the fast magnetosonic mode at high frequencies. Section~3 compares the results from the simulations in Refs.~\cite{Dieckmann2023,Dieckmann2024}. Section~4 summarizes and discusses our findings and explores their possible connection to experimental observations by the MMS mission.

\section{Simulation model}   

All simulations are performed using the PIC code \emph{EPOCH} \cite{Arber2015}, which is based on Esirkepov's algorithm~\cite{Esirkepov2001}. An ambient plasma relevant to laser-plasma experiments is considered. The electrons have a number density $n_{e0}=10^{15}\mathrm{cm}^{-3}$ and temperature $T_e = 1$~keV. The fully ionized nitrogen ions (\textit{Z}=7) have a number density $n_{i0}=n_{e0}/7$ and temperature $T_i = T_e/5$. A magnetic field $\mathbf{B}_0$ with amplitude $B_0=0.85$~T permeates the plasma. The electron plasma frequency and skin depth are $\omega_{pe}={(e^2n_{e0}/\epsilon_0m_e)}^{1/2}$ and $\lambda_e = c/\omega_{pe}$ ($e, m_e, m_i, c, k_B, \epsilon_0, \mu_0$: elementary charge, electron mass, ion mass, light speed, Boltzmann constant, vacuum permittivity and permeability). The correct charge-to-mass ratio for fully ionized nitrogen ions (Z=7) is used, giving them the plasma frequency $\omega_{pi}={(Z^2 e^2 n_{i0}/\epsilon_0 m_i)}^{1/2}$. The electron and ion gyrofrequencies are $\omega_{ce} = eB_0 / m_e$ and $\omega_{ci} = ZeB_0 / m_i$. The lower-hybrid frequency is $\omega_{lh}={((\omega_{ci}\omega_{ce})^{-1}+\omega_{pi}^{-2})}^{-1/2}$. The thermal speeds of the electrons and ions are $v_{te}={(k_B T_e / m_e)}^{1/2}$ and $v_{ti}= {(k_B T_i/m_i)}^{1/2}$.

The ion-acoustic speed $c_s = {(k_B (\gamma_e Z T_e + \gamma_i T_i)/m_i)}^{1/2}$ is the phase and group velocity for ion-acoustic waves with long wavelengths $(\gamma_e = 5/3,\gamma_i = 3$ : adiabatic constants of electrons and ions). The magnetic field with amplitude $B_0$ introduces the Alfv\'en wave with phase- and group velocities $v_A = B_0/{(\mu_0 n_{i0}m_i)}^{1/2}$ and a resonance at $\omega_{ci}$. The fast magnetosonic mode is a  compressive wave, which propagates across the magnetic field and is electromagnetic for frequencies $\omega \ll \omega_{lh}$. Its phase and group velocities become the fast magnetosonic speed $v_{fms}={(c_s^2+v_A^2)}^{1/2}$ for propagation at an angle $\alpha = 90^\circ$ relative to $\mathbf{B}_0$.  Solving the linear dispersion relation for the electrostatic component of the dielectric tensor yields the solution
\begin{equation}
\omega_{ES} = \left (   
3v_{ti}^2 k^2 + \frac{\omega_{pi}^2 (\omega_{ce}^2 + v_{te}^2k^2)}{\omega_{pe}^2 + \omega_{ce}^2+v_{te}^2k^2}
\right )^{1/2}
\label{LHwave}
\end{equation}
for $\omega \ge \omega_{lh}$. Taking into account the full dielectric tensor yields a single wave branch. Table~\ref{table1} lists the values for the frequencies, speeds, and the electron skin depth of the ambient plasma in all simulations.

The fluctuation spectra, computed by a dedicated PIC simulation and connected to the dielectric tensor~\cite{Dieckmann2004}, are used to observe how both modes connect in the ambient plasma with the parameters stated above. The simulation direction is aligned with the x-axis, with $\mathbf{B}_0 = (0,B_0,0)$, and the power spectra $\langle B_y^2  (k,\omega)\rangle$ and $\langle E_x^2  (k,\omega)\rangle$ are sampled. The one-dimensional simulation, used to sample the noise, employs periodic boundary conditions and resolves the box length $475\lambda_e$ along $x$ with $3 \times 10^4$~cells. The simulation covers the time $t_{noise}$, where $\omega_{lh}t_{noise}=44$.

\begin{table}
\begin{tabular}{|r|l|}
\hline
Electron plasma frequency $\omega_{pe}$ : & $1.8 \times 10^{12}\mathrm{s}^{-1}$\\
Ion plasma frequency $\omega_{pi}$ : & $2.9 \times 10^{10}\mathrm{s}^{-1}$\\
Electron gyrofrequency $\omega_{ce}$ : & $1.5 \times 10^{11}\mathrm{s}^{-1}$\\
Ion gyrofrequency $\omega_{ci}$ : & $4.1 \times 10^7 \mathrm{s}^{-1}$\\
Lower-hybrid frequency $\omega_{lh}$ : & $2.4 \times 10^9 \mathrm{s}^{-1}$\\
Electron thermal speed $v_{te}$ : & $1.3 \times 10^7 \mathrm{m/s}$ \\
Ion thermal speed $v_{ti}$ : & $3.7 \times 10^4 \mathrm{m/s}$ \\
Ion-acoustic speed $c_s$ : & $2.9 \times 10^5\mathrm{m/s}$ \\
Alfv\'en speed $v_A$ : & $4.1 \times 10^5\mathrm{m/s}$ \\
Fast magnetosonic speed $v_{fms}$ : & $5.0 \times 10^5\mathrm{m/s}$ \\
Electron skin depth $\lambda_e$ : & $0.17 \mathrm{mm}$ \\
\hline
\end{tabular}
\caption{Frequencies, speeds, and electron skin depth of the ambient plasma.}
\label{table1}
\end{table}

\begin{figure*}
\includegraphics[width=\columnwidth]{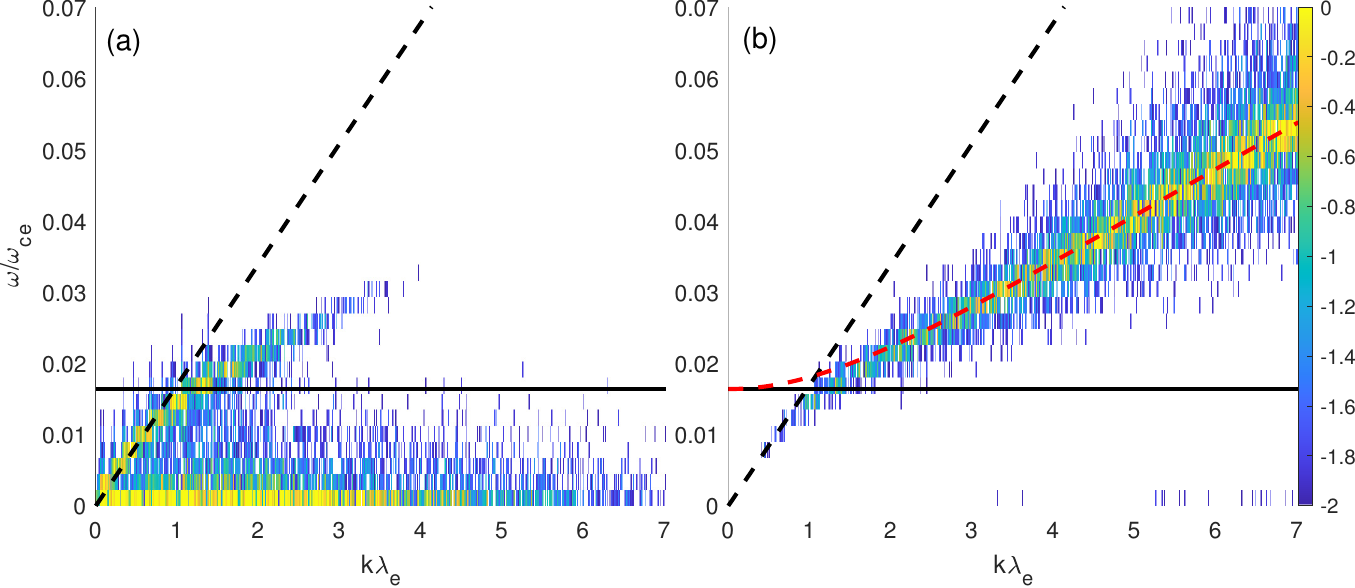}
\caption{Dispersion relation of the charge density wave: Panel~(a) and~(b) show the power spectra of the magnetic fluctuations $\langle B_y^2  (k,\omega) \rangle$ and electrostatic fluctuations $\langle E_x^2  (k,\omega) \rangle$, respectively. The solid black line marks $\omega = \omega_{lh}$ and the dashed black line $\omega = v_{fms}k$. The dashed red curve shows $\omega_{ES}(k)$. The power spectra are normalized to the peak values of the noise propagating on the wave branches and displayed on a 10-logarithmic color scale. The color bar of (b) applies also to (a).}
\label{fig01}
\end{figure*}

Figure~\ref{fig01} shows both power spectra.
Figure~\ref{fig01}(a) reveals magnetic fluctuations with $k\lambda_e < 0.8$ that propagate at the phase velocity $v_{fms}$. Their phase speed decreases for larger values of $k\lambda_e$ as $\omega$ approaches $\omega_{lh}$. On the chosen color scale, the magnetic noise vanishes at $k\lambda_e = 4$. Figure~\ref{fig01}(b) shows fluctuations of $\langle E_x^2  (k,\omega)\rangle$ that follow the dispersion relation of the fast magnetosonic mode. Their low power for $k\lambda_e < 1$ implies that fast magnetosonic modes are predominantly electromagnetic for wavelengths larger than $\lambda_e$. The combined mode becomes predominantly electrostatic for $k\lambda_e > 2$. The change occurs at a wavenumber that is well below $k_g = 2\pi \lambda_e / \lambda_g = 12$, which corresponds to the electron's thermal gyroradius $\lambda_g = v_{te}/\omega_{ce}$. The combined mode becomes dispersive close to $\omega_{lh}$. It is expected that a fast magnetosonic shock steepens until the dominant mode mediating it reaches a wavelength of approximately $\lambda_e$. This will be confirmed by simulations.

\begin{figure*}
\begin{center}
\includegraphics[width=0.8\columnwidth]{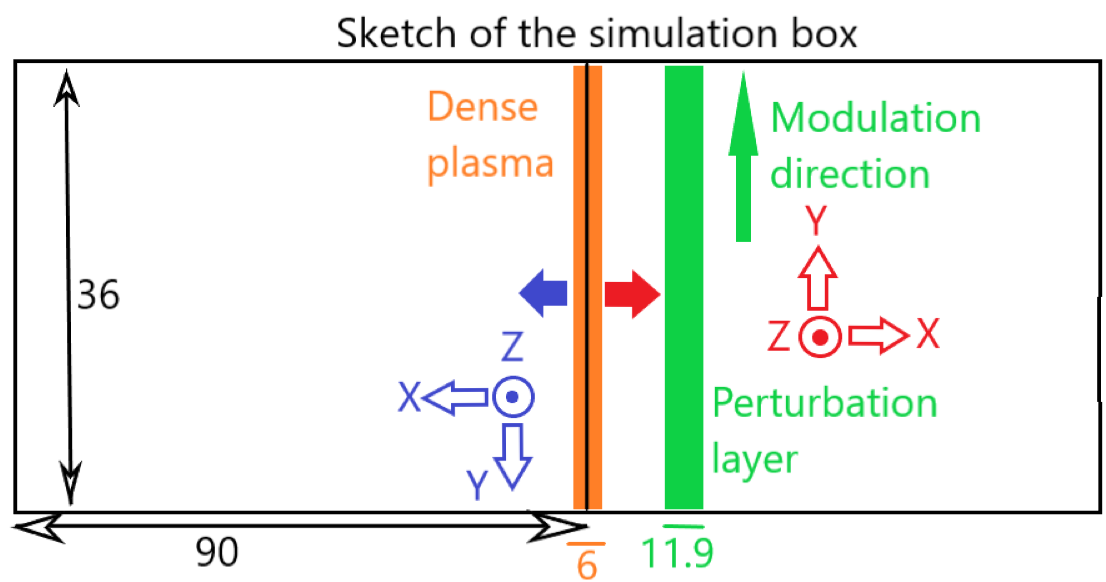}
\end{center}
\caption{The simulation box is subdivided into two halves at the vertical black line, which marks $x=0$. Each half has the with 90$\lambda_e$ along $x$ as marked by the horizontal double arrow. The vertical width 36$\lambda_e$ of the box is shown by the vertical double arrow. A layer of dense plasma is placed in the center of the simulation box. Its width is 6$\lambda_e$ and it is cut in half by $x=0$. The dense plasma is surrounded by ambient plasma. Thermal pressure causes the dense plasma to expand in the directions denoted by the blue and red solid arrows. The blast waves propagate to increasing $x$ in their respective right-handed coordinate system. The blast wave in the blue coordinate system expands into a spatially uniform ambient plasma and serves as the reference shock. The blast wave in the red coordinate system eventually crosses the perturbation layer, where the number density of mobile ions varies sinusoidally along the modulation direction $y$. The width of the perturbation layer is 11.9$\lambda_e$ and it covers $8.9\le x / \lambda_e \le 20.8$.}
\label{fig02}
\end{figure*}

In the next section, simulations are performed in two spatial dimensions, resolving the $x,y$-plane and employing periodic boundary conditions in all directions. The box is subdivided into two halves as shown in Fig.~\ref{fig02}. 
The simulation box is filled with ambient plasma with the initial conditions listed above, except in the intervals marked in orange and green. The box length $L_y=36\lambda_e$ along $y$ is resolved by 1800 grid cells. Each of the two domains, marked by the blue and red coordinate systems, has the length 90$\lambda_e$ along $x$, resolved by 4500 grid cells, respectively. The dense plasma has a width 6$\lambda_e$ along $x$ and is distributed equally over both box halves, covering $0 \le x/\lambda_e \le 3$ in each half-space. The number densities of electrons and ions are 60 times those of the ambient plasma, and their temperatures are $1.5T_e$ and $T_i$, respectively.
 
The thermal pressure of the dense plasma causes it to expand into the ambient plasma, forming a blast wave, as discussed in detail in Ref.~\cite{Ly2023}. The perturbation layer covers $8.9 \le x/\lambda_e \le 20.8$ in the red coordinate system. In this layer, the number density of mobile ions varies as $n_{i0,mob}/n_{i0}= (0.7 + 0.3\sin{(2\pi y / L_y)})$, while the electron density remains $n_{e0}$. Since the electric field is set to $\mathbf{E}=0$ at the start of the simulation, the net charge of the mobile plasma is compensated by an immobile positive charge density $n_{i0}-n_{i0,mob}$, which acts as a grating. This setup is employed by both simulations discussed in the next section. Both differ only in the direction of $\mathbf{B}_0$.

\section{Results}

In this section, the results of the two two-dimensional PIC simulations, which were discussed separately in~\cite{Dieckmann2023,Dieckmann2024}, are compared directly. Both simulations model a plasma with the setup discussed in Section~2. The simulations are labeled according to the initial magnetic field direction. Simulation~Y refers to the 2D simulation in~\cite{Dieckmann2023}, which set the magnetic field direction to $\mathbf{B}_0 \parallel y$. Simulation~Z~\cite{Dieckmann2024} aligned the magnetic field $\mathbf{B}_0$ with the numerically unresolved $z$ direction.

\begin{figure*}
\includegraphics[width=\columnwidth]{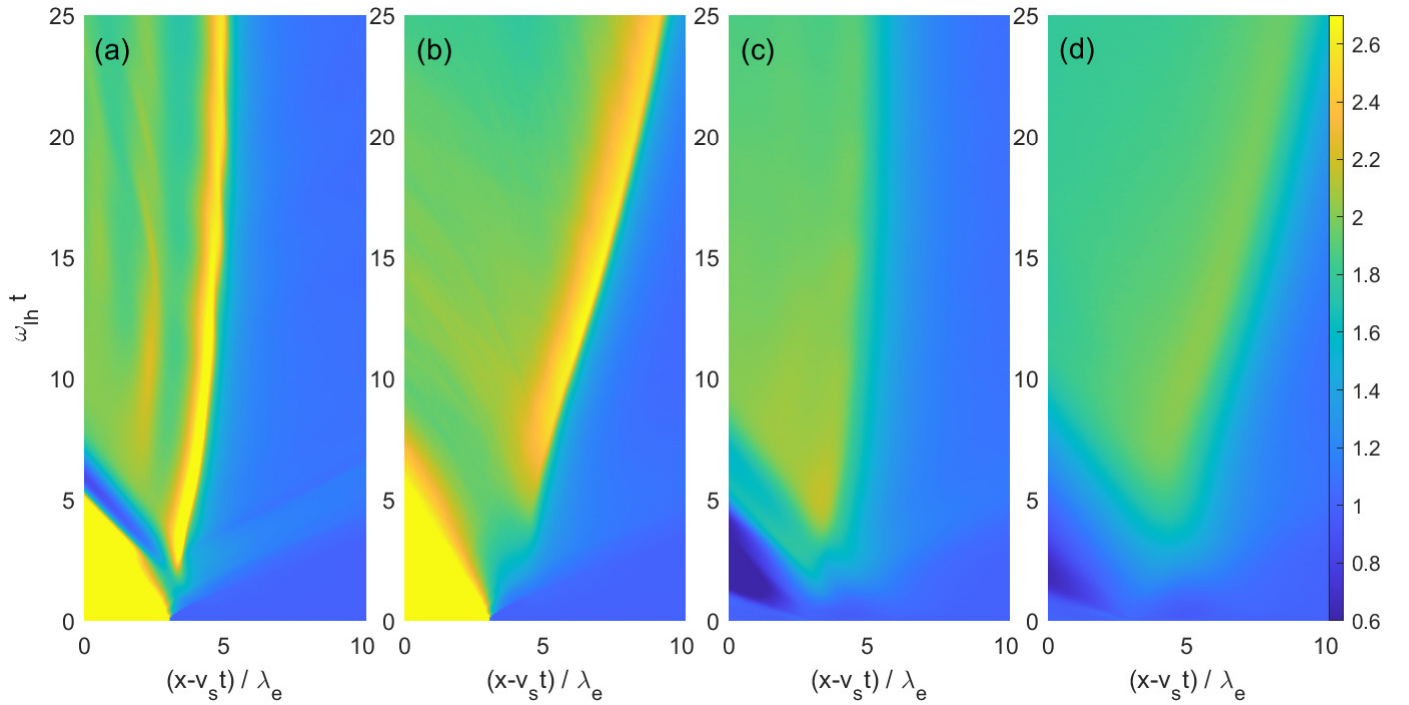}
\caption{Box-averaged ion density and magnetic field amplitude along the direction of $\mathbf{B}_0$ of the reference shocks: Panels~(a,~b) show the ion densities $n_i(x)/n_{i0}$ computed by simulations~Y and~Z, respectively. Panel~(c) shows $B_y(x)/B_0$ computed by simulation~Y, and panel~(d) shows $B_z(x)/B_0$ computed by simulation~Z. The linear color scale of~(d) applies to all panels, and $v_s = 1.6v_{fms}$ is the speed of the moving window.}
\label{fig03}
\end{figure*}

Figure~\ref{fig03} compares the reference shocks in simulations~Y and~Z, which propagate into the left half of the simulation box shown in Fig.~\ref{fig02}. Ion densities and magnetic fields are averaged over the interval $0 \le y/\lambda_e \le 36$. Moving windows, which propagate at the speed $v_s = 1.6v_{fms}$ in the direction of increasing $x$ in the blue coordinate system, were used. 
Figures~\ref{fig03}(a,~b) compare the ion densities in simulations~Y and~Z. The reference shock in simulation~Z is faster by about $0.15 v_{fms}$ compared to that in simulation~Y. Additionally, ripples can be observed in the ion density behind the reference shock in simulation~Y, separated by a distance of about 2$\lambda_e$ from the shock front. These ripples correspond to the density maxima of fast magnetosonic waves that build up the subcritical shock. According to Fig.~\ref{fig01}, their wavenumber $k\lambda_e \approx \pi$ indicates that they are predominantly electrostatic. No such ion density ripples are observed in simulation~Z. Both reference shocks amplify the magnetic field to a similar downstream value, but the magnetic overshoot is larger in simulation~Z. The amplification of the magnetic field by both shocks is comparable to the compression of ion density, which is consistent with the frozen-in theorem.

In Ref.~\cite{Dieckmann2023}, one-dimensional simulations were conducted, where the magnetic field was oriented perpendicularly to the simulation direction, to investigate how a fast magnetosonic shock responds to changes in the number density of mobile ions in the perturbation layer shown in Fig.~\ref{fig02}. A comparison of the shocks moving through a perturbation layer with mobile ion number densities of $0.4n_{i0}$ and $n_{i0}$ showed that both shocks remained stable, but the one passing through the diluted ions fell behind the other. Upon exiting the perturbation layer, the lag between the shocks amounted to approximately $\lambda_e$.

\begin{figure*}
\includegraphics[width=\columnwidth]{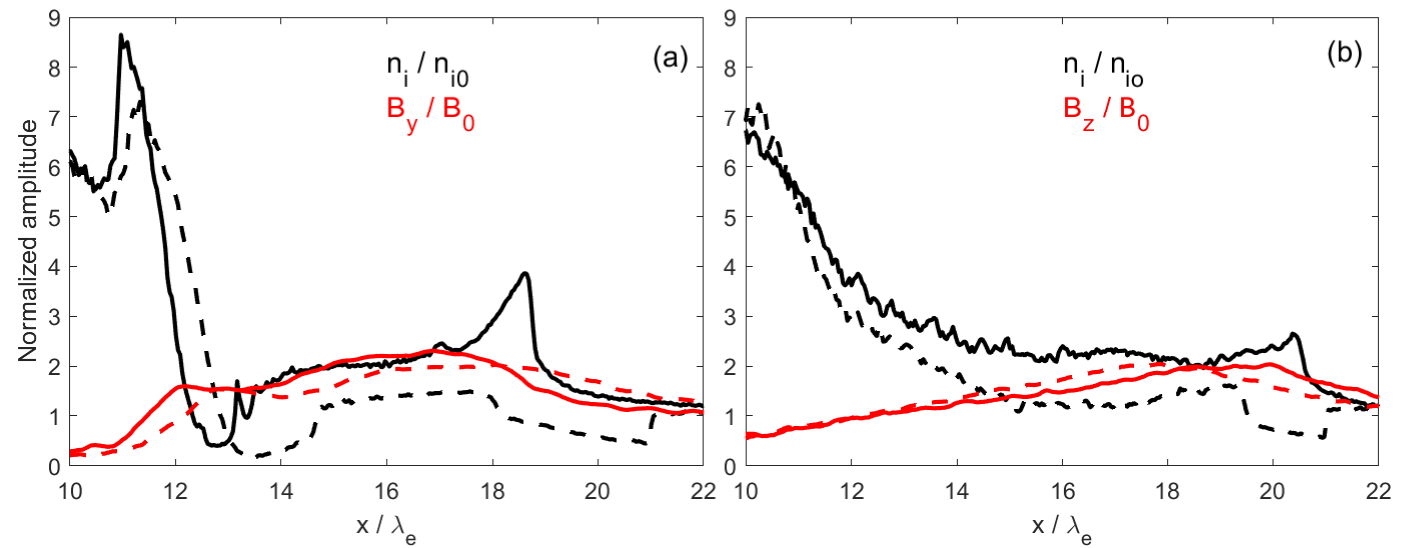}
\caption{Ion density and magnetic field at $\omega_{lh}t=7.9$: Panel~(a) plots $n_i(x)/n_{i0}$ and $B_y(x)/B_0$ computed by simulation~Y, which have been averaged over $8.6 \le y/\lambda_e \le 9.4$ (solid curves) and over $26.6 \le y/\lambda_e \le 27.4$ (dashed curves). Panel~(b) plots $n_i(x)/n_{i0}$ and $B_z(x)/B_0$ computed by simulation~Z, which have been averaged over $8.6 \le y/\lambda_e \le 9.4$ (solid curves) and over $26.6 \le y/\lambda_e \le 27.4$ (dashed curves).}
\label{fig04}
\end{figure*}

Similar lags are also observed in simulation~Y in Fig.~\ref{fig04}(a) and simulation~Z in Fig.~\ref{fig04}(b) when the shocks nearly reach the end of the perturbation layer. The mobile ion density in the perturbation layer is highest in the slice $y\approx 9\lambda_e$ and lowest in the slice $y\approx 27\lambda_e$.
Figure~\ref{fig04}(a) shows a rapid decrease in ion density at $x\approx 12\lambda_e$, coinciding with an increase in the magnetic amplitude, indicating the presence of tangential discontinuities in both slices. A tangential discontinuity separates the dense, weakly magnetized plasma of the blast wave from the diluted, magnetized ambient plasma. The ion density $n_i(x)/n_{i0}$ to the right of the tangential discontinuity in Fig.~\ref{fig04}(a) decreases well below 1, caused by an ion phase space vortex. The ion density difference in the interval $15 \le x/\lambda_e \le 18$ is comparable to the difference of 0.6 in $n_{i0,mob}/n_{i0}$ between $y\approx 9\lambda_e$ and $27\lambda_e$. The density overshoot of the shock in the slice $y\approx 9\lambda_e$ is located at $x\approx 18.5\lambda_e$. As the magnetic field is amplified by the shock crossing, the shock is mediated by the fast magnetosonic mode. In the slice $y\approx 27\lambda_e$, an ion density change is observed at $x\approx 18\lambda_e$, but neither a density overshoot nor a rapid change in magnetic amplitude is observed at its position. 

The shocks have propagated farther in simulation~Z than in simulation~Y. While the density overshoot of the shock at $x=20.5\lambda_e$ in Fig~\ref{fig04}(b) at $y\approx 9\lambda_e$ is not as large as the one in Fig.~\ref{fig04}(a) at $x=18.5\lambda_e$, the density behind the shock is twice that ahead of it, as expected from shocks that heat ions only along one direction~\cite{Livadiotis2015}. At low $x$ in Fig.~\ref{fig04}(b), the ion density decreases slowly, and the magnetic amplitude increases gradually with $x$. The broadening of the $x$-interval, compared to that in Fig.~\ref{fig04}(a), where these changes occur, can be attributed at least partially to drift instabilities. A change in the magnetic field amplitude is caused by a current density $\mathbf{J}\neq 0$, due to electrons drifting relative to ions. In simulation~Y, the change in the amplitude of $B_y$ is sustained by a current along the numerically unresolved z-direction, preventing the growth of either the electron-cyclotron drift instability~\cite{Forslund1970b,Forslund1972} or the lower-hybrid drift instability~\cite{Davidson1977,Huba1978,Drake1983}. If the electric current is not dissipated, rapid spatial variations in the magnetic field amplitude are possible. Simulation~Z resolves the direction along which the electrons drift, and instabilities dissipate the electric current. Figure~\ref{fig04} shows that the dissipation of the electric current in simulation~Z also makes it more difficult for the blast wave to expel the magnetic field as it expands to increasing $x$.

\begin{figure*}
\includegraphics[width=\columnwidth]{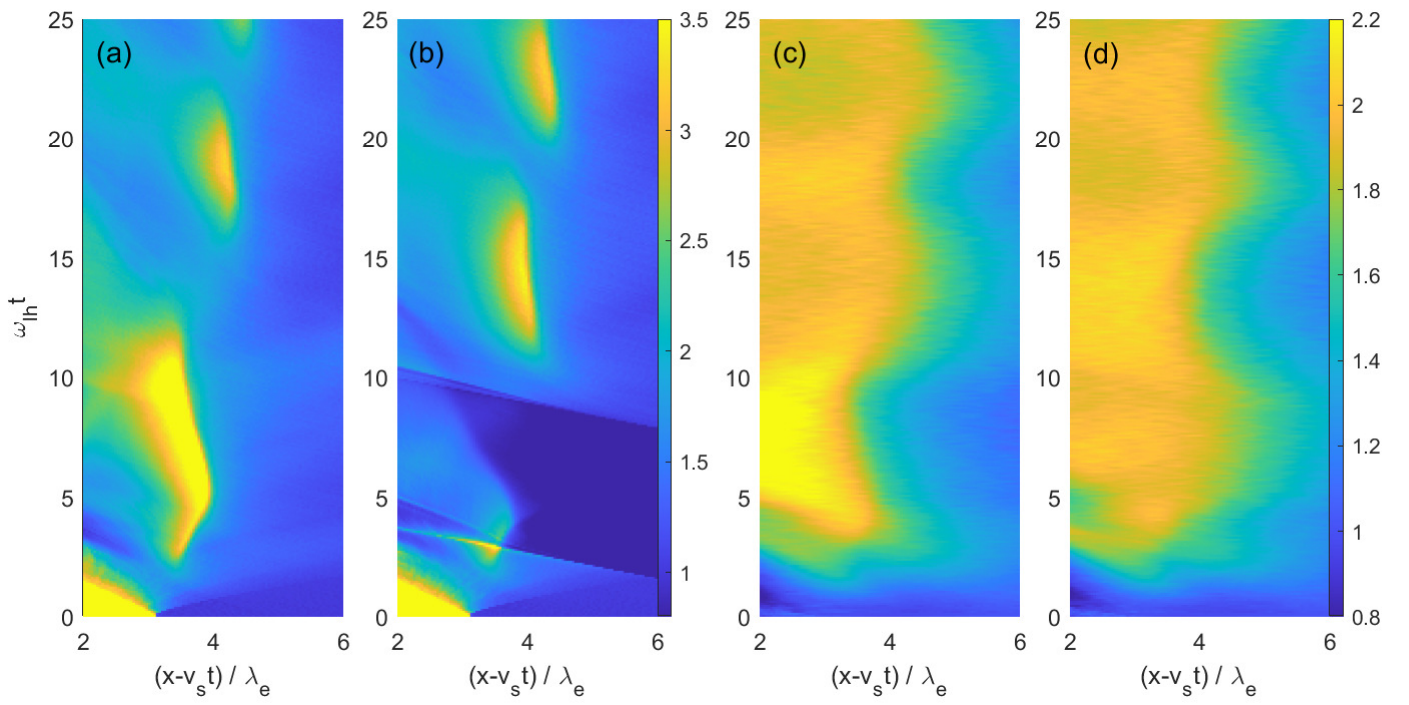}
\caption{Ion density and magnetic field amplitudes of the perturbed shock in simulation~Y: Panels~(a,~b) show the normalized ion densities $n_i(x)/n_{i0}$ averaged over the intervals $8.6 \le y/\lambda_e \le 9.4$ and $26.6 \le y/\lambda_e \le 27.4$, respectively. Panels~(c,~d) show $B_y(x)/B_0$ averaged over the same intervals. Panels~(a,~b) use the linear color scale of (b), and panels~(c,~d) use the linear color scale of~(d). The moving window has the speed $v_s = 1.6v_{fms}$.}
\label{fig05}
\end{figure*}

The ion densities and amplitudes of $B_y$ for the perturbed shock in simulation~Y are examined in Fig.~\ref{fig05}. Both quantities are averaged over an interval with a width of 0.8$\lambda_e$, centered on the coordinate values $y=9\lambda_e$ and $y=27\lambda_e$. These are displayed in a moving window that travels with the speed $v_s = 1.6 v_{fms}$ to increasing $x$ in the red coordinate system shown in Fig.~\ref{fig02}. 
A comparison of Figs.~\ref{fig05}(a,~b) reveals ion density oscillations with a phase shift of $180^\circ$ after the shock has exited the perturbation layer. The shock is initially planar before entering the perturbation layer, with its normalized density reaching 3.5 at the overshoot near $x-v_st = 3.5\lambda_e$ and $\omega_{lh}t=3$. During $3 \le \omega_{lh}t \le 5$, the shock in the slice $y\approx 9\lambda_e$ in Fig.~\ref{fig05}(a) and the reference shock propagate at similar speeds. After $\omega_{lh}t=5$, its speed decreases, coinciding with the collapse of the shock in the slice $y\approx 27\lambda_e$. A comparison of Figs.~\ref{fig05}(a,~c) for $5 \le \omega_{lh} t \le 10$ shows that the magnetic field changes are correlated with those of the ion density. The structure receding from $x-v_st = 4\lambda_e$ to $x-v_st =3.5\lambda_e$ during this time is a fast magnetosonic shock that is slower than the reference shock. In contrast, the magnetic field in Fig.~\ref{fig05}(d) within the slice $y\approx 27\lambda_e$ expands outward during $5 \le \omega_{lh}t \le 10$, and its front at $x-v_st \approx 4\lambda_e$ does not visibly correlate with a change in $n_i$ in Fig.~\ref{fig05}(b). Once the shock boundary leaves the perturbation layer, it begins to oscillate with a period of approximately 10/$\omega_{lh}$. Each of the density peaks in Figs.~\ref{fig05}(a,~b) coincides with a magnetic field confined downstream of the density overshoot. Density minima correlate with a magnetic field that bulges into the upstream plasma. Although the fully developed fast magnetosonic shocks propagate at a speed below $v_s$ and thus below that of the reference shock the oscillating boundary moves faster since the slope of a line connecting the three density maxima after $\omega_{lh}t=10$ moves to increasing $x-v_st$.   

\begin{figure*}
\includegraphics[width=\columnwidth]{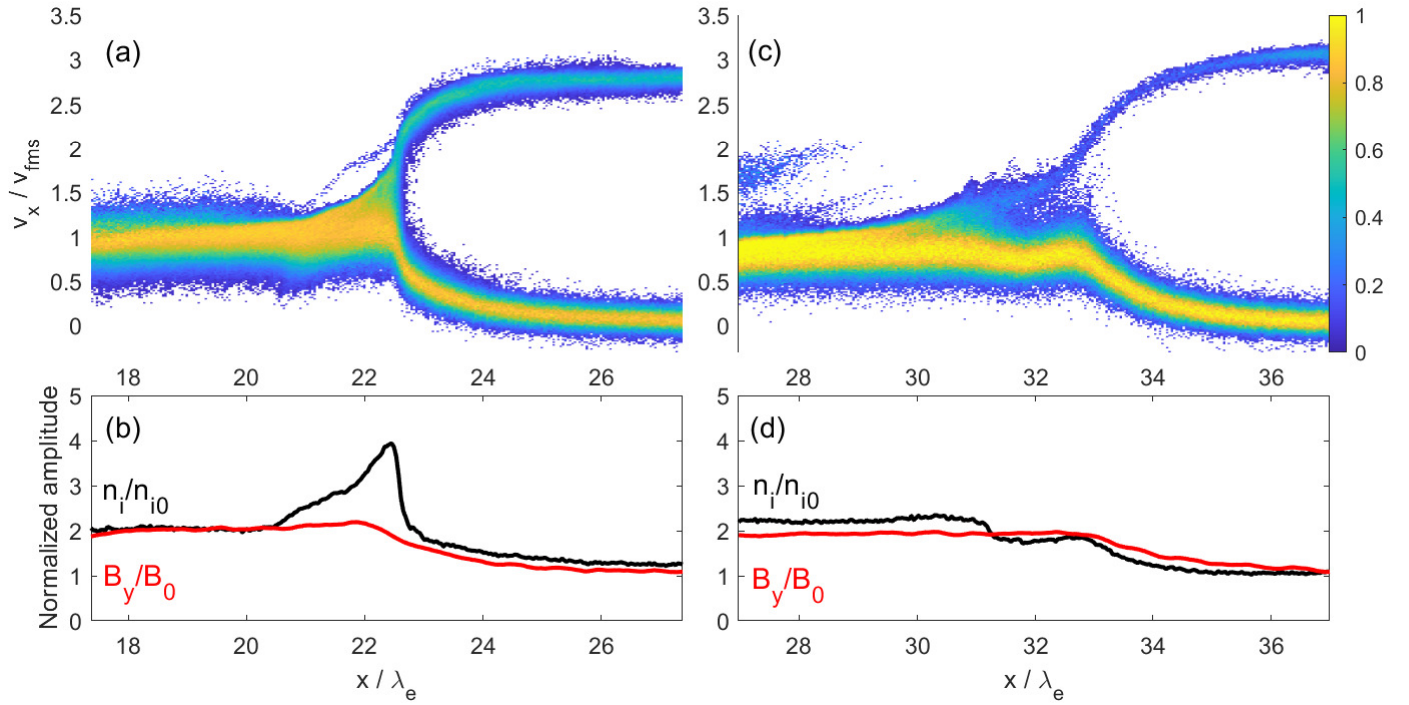}
\caption{Distributions $f_i(x,v_x)^{1/2}$, $n_i(x)/n_{i0}$ and $B_y(x)/B_0$ in the slice $y\approx 9\lambda_e$ at times 10/$\omega_{lh}$ (left column) and 15/$\omega_{lh}$ (right column) for the perturbed shock in simulation~Y. The ion phase space densities are normalized to the peak value of the ambient plasma. The square root of the values at both times is displayed on the same linear color scale clamped to 1. The color bar of~(c) applies also to~(a). All displayed quantities have been averaged over $8.6 \le y/\lambda_e \le 9.4$.}
\label{fig06}
\end{figure*}

Figure~\ref{fig06} compares the ion phase space density distributions $f_i(x,v_x)$ with the distributions of $n_i(x)/n_{i0}$ and $B_y(x)/B_0$ in the slice $y\approx 9\lambda_e$, corresponding to Figs.~\ref{fig05}(a,~c), at times 10/$\omega_{lh}$ and 15/$\omega_{lh}$ in the red coordinate system in Fig.~\ref{fig02}.
Figs.~\ref{fig06}(a,~b) show the distributions at time 10/$\omega_{lh}$, when the shock is fully developed. It is located at $x = 22.5\lambda_e$, with a peak density overshoot value of 4. The magnetic field amplitude and ion density behind the shock are twice their upstream values. The electric potential difference between the overshoot and upstream plasma is large, and a significant fraction of inflowing upstream ions are reflected. Shortly after the time, the shock collapses, as shown in Fig.~\ref{fig05}(a). We find the remnant of the overshoot at $x\approx 31\lambda_e$ in Fig.~\ref{fig06}(c). This ion accumulation is still connected to the shock-reflected ion beam, but it no longer injects ions into it. A new wave is growing at $x\approx 32.5\lambda_e$, which will develop into a new fast magnetosonic shock. Figure~\ref{fig06}(d) shows that although the density ratio between downstream and upsteam plasma is still about 2, the density overshoot has disappeared. The magnetic field has expanded, and its amplitude change surpasses the ion density change up to $x=36\lambda_e$. This magnetic field advances in $x$, trapping electrons and creating an electric current. The electric field induced by this current accelerates the ions within the interval $33 \le x / \lambda_e \le 37$, building up the wave. The structure in Figs.~\ref{fig06}(c,~d), which separates the downstream and upstream regions, is thus a magnetic piston rather than a shock. Movie~1 shows the time evolution of the quantities displayed in Fig.~\ref{fig06}(a,~b) in the slice $y\approx 9\lambda_e$ over $0 \le \omega_{lh} t \le 25$, revealing how the shock changes into a piston and back.  

\begin{figure*}
\includegraphics[width=\columnwidth]{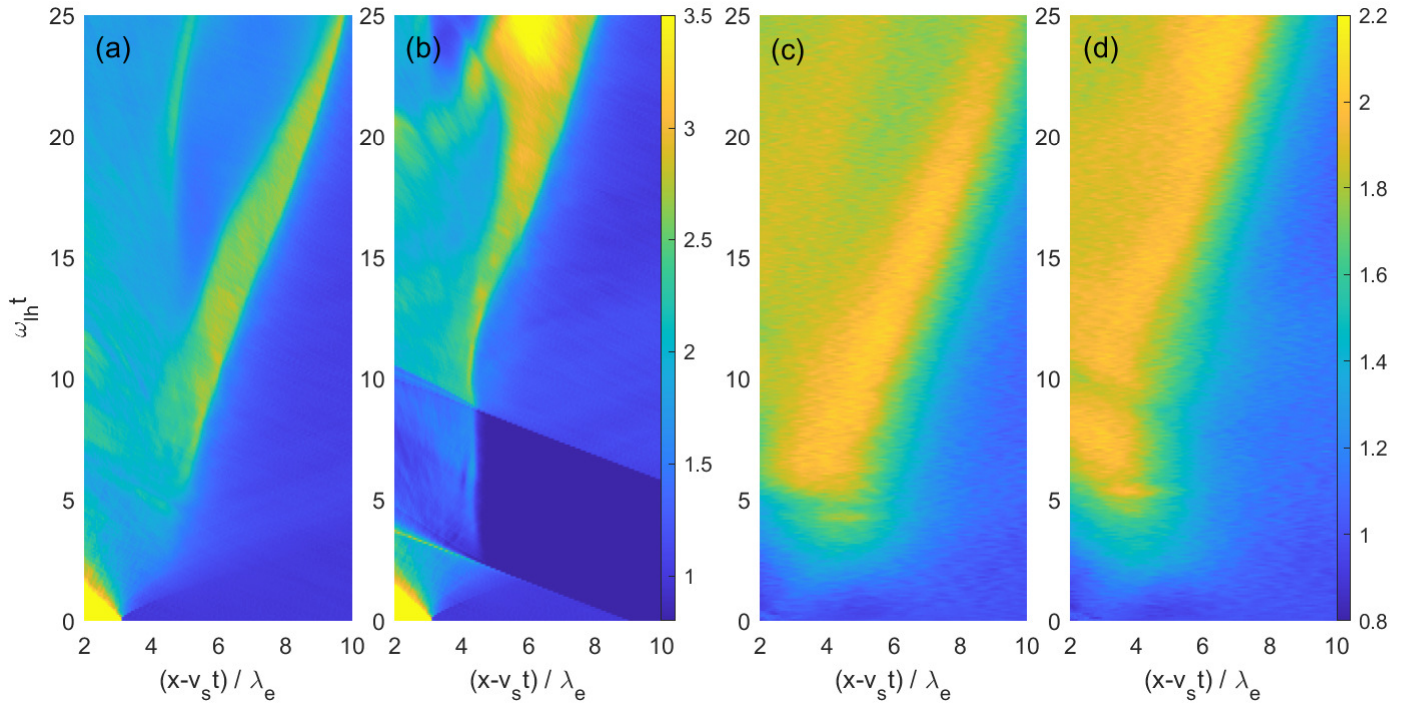}
\caption{Ion density and magnetic field amplitudes of the perturbed shock in simulation~Z: Panels~(a,~b) show the normalized ion densities $n_i(x)/n_{i0}$ averaged over intervals $8.6 \le y/\lambda_e \le 9.4$ and $26.6 \le y/\lambda_e \le 27.4$, respectively. Both panels use the linear color scale of~(b).
Panels~(c,~d) show $B_z(x)/B_0$ averaged over the same intervals. They use the linear color scale of (d). The moving window has the speed $v_s = 1.6v_{fms}$.}
\label{fig07}
\end{figure*}

Figure~\ref{fig07} shows the time-evolution of ion densities and $B_z/B_0$ amplitudes along the slices $y\approx 9\lambda_e$ and $y\approx 27\lambda_e$, calculated by simulation~Z, where the ambient magnetic field aligns with the numerically unresolved $z$-direction. 
Neither the ion density nor the magnetic field oscillates over time; both fronts of the compressed magnetic field align with those of the density. This alignment persists even as the shock traverses the perturbation layer in Fig.~\ref{fig07}(b). Passing through the perturbation layer slows the shock, which then accelerates after leaving the perturbation layer and reaches the same speed as the shock in the slice $y\approx 9\lambda_e$ at $\omega_{lh}t\approx$ 15. However, it cannot catch up with the shock in the $y\approx 9\lambda_e$ slice within the simulation time. A slow decrease in the separation between the shocks in both y-slices does indicate, though, that the deformed shock is stable. At later times, the density overshoot in Fig.~\ref{fig07}(a) becomes thinner while ions accumulate behind the shock front in Fig.~\ref{fig07}(b). The thickness $\sim 2\lambda_e$ is comparable to that of waves driven by the lower-hybrid drift instability observed in Ref.~\cite{Dieckmann2024}. Another contribution comes from ions deflected by adjacent tilted shock fronts towards the trailing shock near $y\approx 27\lambda_e$. Ions slow down along the shock normal when they cross the shock but their velocity component along the shock plane remains unchanged causing a drift along it.  

\section{Discussion}

The reaction of subcritical fast magnetosonic shocks to perturbations of the shock boundary has been compared. In the two-dimensional simulation~Y, which is discussed in depth in Ref.~\cite{Dieckmann2023}, the background magnetic field was aligned with $y$, while in simulation~Z discussed in Ref.~\cite{Dieckmann2024}, it was aligned with the numerically unresolved direction $z$ perpendicular to the simulation box. It was found in these previous works that the shock front is stable in both simulations, as the oscillations do not continue to grow once the shock front has exited the perturbation layer. The shock front oscillated in simulation~Y at a frequency just below the lower-hybrid frequency and the oscillations were weakly damped. The oscillations of the ion density and magnetic field were shifted by 180$^\circ$. In simulation~Z the perturbation was non-oscillatory and weakly damped, with the magnetic field near the shock closely following that of the ion density.

It was shown in Ref.~\cite{Dieckmann2024} that electron cyclotron drift instabilities and lower-hybrid drift instabilities develop in simulation~Z but are unresolved in simulation~Y. These drift instabilities dissipate the electronic current that maintains spatial changes in the magnetic field near the shock or the trailing tangential discontinuity. This allows the magnetic field to diffuse more easily into the plasma in simulation~Z, unlike in simulation~Y, which may influence the shock expansion speed in simulation~Z.

Here we compared the results of both simulations in more detail. The speed of the reference shock in simulation~Z was indeed higher than that in simulation~Y. The key finding of this comparison was that the oscillations of the perturbed shock in simulation~Y are caused by a periodic collapse and reformation of the shock. The boundary changes periodically from a fast magnetosonic shock, where the magnetic field change follows that of the density change, into a magnetic piston, where the magnetic field bulges into the upstream. The oscillation frequency just below the lower-hybrid frequency reported in Ref.~\cite{Dieckmann2023} is likely to be a consequence of the time it takes the magnetic piston to grow a new wave and turn it into a shock.

In a 3D setting, it is expected that shock oscillations would occur along, but not perpendicular to the background magnetic field, as observed by the MMS mission~\cite{Johlander2016}. However, there are key differences between these scenarios. Solar wind consists primarily of protons, while fully ionized nitrogen, often used in laser-plasma experiments, was modeled here. If both ion species are given the same temperature, the thermal velocity of the protons is significantly higher than that of nitrogen, and thermal effects such as damping may be stronger. Additionally, the shock observed by the MMS mission was supercritical and, thus, not stationary, reforming on time scales comparable to the inverse ion gyrofrequency. Here, a subcritical shock was considered. It is interesting to note in this context that oscillations of the boundary of a supercritical shocks with frequencies comparable to the ion-acoustic frequency were observed in the PIC simulation in Ref.~\cite{MTSI}. The mechanism of their generation was not identified and may be related to the one discussed here. Finally, the MMS mission detected Alfv\'enic waves with frequencies much lower than the oscillations investigated here. Alfv\'enic waves also have wavelengths that far exceed the box size of our simulations. 

There is, however, one aspect the oscillations in the simulations presented here and those observed by the MMS mission have in common. In the simulation, oscillations were caused by a periodic change between a shock and a magnetic piston at the boundary. Simulation~Y suggests a loop involving the following steps. (1)  A fast magnetosonic shock (the old shock) causes a rapid change in the magnetic amplitude near the density overshoot, and both are spatially correlated. In the case of the piston, the plasma cannot confine the magnetic field in the downstream region, so it bulges into the upstream domain, creating a sinusoidal deformation in the magnetic field with a 180$^\circ$ phase shift relative to that of the density~\cite{Dieckmann2023}. (2) The magnetic piston drags trapped electrons across the upstream ions, which results in an electric current. This current induces an electric field, which causes the growth of a wave at the piston's front. Eventually, this wave changes into a shock. (3) Once this shock formed, it accelerates upstream ions to a higher downstream velocity and injects some into the shock-reflected ion beam. The increased ram pressure pushes the shock toward the downstream plasma. Since the new shock and the old shock are connected via the magnetic field, the magnetic field lines move relative to the old shock. This movement alters the dispersive properties of the fast magnetosonic modes that form the old shock, causing it to collapse as the magnetic field bulges into the upstream plasma. The oscillation period is set by the time required for a new shock wave to develop. Consequently, the shock oscillations in simulation~Y result from a periodic collapse and reformation of the shock front. Supercritical shocks also undergo cyclic reformation, albeit over longer time scales. 

Since the magnetic field is deformed by the shock reformation, it can potentially couple to magnetowaves. A deformation perpendicular to the magnetic field direction at a frequency below the ion gyrofrequency could couple to propagating Alfv\'en modes. Magnetic field oscillations close to the lower-hybrid frequency, such as those observed here, cannot couple to Alfv\'en waves but may couple to Whistler waves, which can reach much higher frequencies. Whistler waves are not necessarily propagating along the magnetic field. Oblique Whistlers can grow ahead of the shock due to the modified two-stream instability between the shock-reflected ion beam or the inflowing upstream ions and the upstream electrons. Oblique Whistler waves grew ahead of the supercritical perpendicular fast magnetosonic shock in Ref.~\cite{MTSI}. We did not observe this instability in our simulations but it may grow over longer spatiotemporal scales and at angles relative to $\mathbf{B}_0$, which are not resolved by the simulations discussed here. These aspects will be left for future work.

\section*{Acknowledgements} The simulations were performed on resources provided by the National Academic Infrastructure for Supercomputing in Sweden (NAISS) at the National Supercomputer Centre partially funded by the Swedish Research Council through grant agreement no. 2022-06725 and on the centers of the Grand Equipement National de Calcul Intensif (GENCI) under grant number A0090406960. The first author also acknowledges financial support from a visiting fellowship of the Centre de Recherche Astrophysique de Lyon. A.B. acknowledges support from the Ministerio de Economía y Competitividad of Spain (Grant No. PID2021-125550OBI00).

\section*{Data availability statement} All data that support the findings of this study are included within the article (and any supplementary files).

\section*{Conflict of interest} The authors declare that they have no conflict of interest.

\end{document}